\documentclass[10pt,journal]{IEEEtran}

\usepackage[T1]{fontenc}
\usepackage{amsmath,amssymb,amsfonts}
\usepackage{newtxtext,newtxmath} % Times-like text & math (IEEEtran-friendly)
\usepackage{cite}
\usepackage{algorithmic}
\usepackage{graphicx}
\usepackage{xcolor}
\usepackage{listings}
\usepackage{array}
\usepackage{xspace}
\usepackage{url}
\usepackage[hidelinks]{hyperref}
\usepackage{caption}

\lstdefinelanguage{json}{
  basicstyle=\ttfamily\small,
  numbers=left, numberstyle=\tiny, stepnumber=1, numbersep=4pt,
  showstringspaces=false,
  morestring=[b]",
  morecomment=[l]{//}
}
\newcommand{\agenticjwt}{Agentic JWT (A-JWT)\xspace}

\title{Agentic JWT: A Secure Delegation Protocol for Autonomous AI Agents}

\author{Abhishek~Goswami,~\IEEEmembership{Senior~Member,~IEEE}%
\thanks{Correspondence: \href{mailto:agoswam1@uchicago.edu}{agoswam1@uchicago.edu}.}%
}

\begin{document}
\maketitle

% ----- Abstract & (optional) keywords -----
% Ensure abstract.tex uses \begin{abstract} ... \end{abstract}
% If you keep keywords, use \begin{IEEEkeywords} ... \end{IEEEkeywords}
\begin{abstract}\label{sec:abstract}
Autonomous LLM agents can issue thousands of API calls per hour without human oversight.  
OAuth 2.0 assumes deterministic clients, but in agentic settings stochastic reasoning, 
prompt injection, or multi-agent orchestration can silently expand privileges.

This paper describes \agenticjwt, a dual-faceted token design that binds each agent action to a 
cryptographically verifiable user intent and optionally to a workflow step. A-JWT carries an 
agent's identity as a one-way checksum hash derived from its prompt, tools and configuration and a chained 
delegation assertion to prove which downstream agent may execute a given task. The design also uses per-agent 
proof-of-possession keys to prevent replay and in-process impersonation. The paper introduces a 
new unique authorization grant called `agent\_checksum` and adds a lightweight client shim library that 
self-verifies code at run time, mints intent tokens, tracks workflow steps and derives keys thus enabling 
secure agent identity and separation even within a single process.  

We illustrate a comprehensive threat model for agentic applications, implement a Python proof-of-concept, 
and show functional blocking of scope-violating requests, replay, impersonation, and prompt-injection pathways 
with sub-millisecond overhead on commodity hardware. The design aligns with ongoing OAuth agent discussions and 
offers a drop-in path toward zero-trust guarantees for agentic applications. A comprehensive performance and security 
evaluation with experimental results will appear in our forthcoming journal submission.
    
\end{abstract}

% ----- Main content -----
\section{Introduction}
\label{sec:introduction}
\textbf{AI Agents} are not a theoretical phenomenon anymore. Large enterprises now use 
AI agents \cite{msft_fq24_q3}, to possibly execute millions of API calls per hour. Major 
cloud LLMs now serve hundreds of millions of API requests per day, for example Baidu’s 
ERNIE handles approximately 200 M daily queries, providing the raw horsepower that agent 
frameworks build on \cite{baidu_fy24_q1}, yet those calls still ride on OAuth tokens 
designed for deterministic clients.

A quick peek into the scale of operations and future trends would reveal that the volume of 
AI Agent activity has grown dramatically, underscoring their operational impact. Baidu’s 
large volume of API calls per day has seen a 4 fold increase in just a few months \cite{baidu_fy24_q1}. 
A recent cloud survey found OpenAI/Azure AI services are used in 67\% of cloud deployments, 
alongside a rise in self-hosted AI models across 75\% of organizations \cite{wiz_ai_cloud_2025}. 
Deloitte projects that a quarter of companies using Generative AI will pilot Agentic AI by 
2025 and this will grow to 50\% by 2027 \cite{deloitte_tmt_2025_agents}. Meanwhile, Gartner 
predicts that by 2028, 33\% of enterprise applications will include Agentic AI \cite{gartner_agentic_ai_2024}. 

Function calling (including API calling) driven by LLM instructions is the fundamental unit 
of work for Autonomous AI agents and exposes API servers and applications to unique security 
threats never anticipated by any of the existing AuthN / AuthZ models. The boundaries of 
Zero Trust architecture can be potentially broken by these nondeterministic agent calls. 
The best API protection methodology today, namely OAuth 2.0, is based on the understanding 
that the caller of an API accurately represents the runtime intent of the user (Resource Owner 
in OAuth terminology) on whose behalf the call is taking place. OAuth 2.0 \cite{rfc6749} 
contains clear understanding that a client's request embodies the resource owner’s intent. In the 
standard flow, “the authorization server issues an access token to the client with the 
approval of the resource owner” (RFC 6749 §1.4) \cite{rfc6749}. The client attaches that 
token to each API call; the resource server validates it and “executes the request on behalf 
of the resource owner” (RFC 6749 §1.1). For the token’s lifetime, the resource server 
therefore treats the client and the user as indistinguishable \cite{rfc6749}

However, in the Agentic AI world all actions (including API calls) are fueled by the decisions taken by an LLM (or a Generative AI model). It is no more a fixed representation of a user’s legitimate / registered intent. For instance, An LLM can be prompted (either inadvertently or on purpose) to generate instructions to use AI agents with elevated privileges, especially when AI Agents are not cryptographically isolated as separate identities. So far, existing API authorization standards do not provide any mechanism for cryptographically distinguishing between execution and intent and rightly so because they were never designed for this purpose. For example, OAuth 2.0 standards like JWT use ‘claims’ to make assertions about a given client. The client is supposed to have registered with an IDP (Identity Provider) that supports JWT tokens. JWTs are commonly used as bearer access tokens, any party that possesses the token is accepted as acting for the resource owner. RFC 6750 defines a bearer token as “a security token with the property that any party in possession of the token can use it” (§1) \cite{rfc6750}. JSON Web Token itself is a self-contained, compact representation of claims (RFC 7519 §2) that can be “used for client authentication and authorization” without requiring additional secrets \cite{rfc7519}. In practice, therefore, whoever carries a signed JWT is treated as a true proxy for the user’s intent for as long as the token remains valid.

This assumption breaks in the agentic-AI world, where a large-language model can dynamically generate multi-step action plans through chain-of-thought reasoning, selecting agentic tools and parameters on the fly rather than following a fixed user workflow \cite{wei_cot_2022,yao_treeofthought_2023}. This causes a separation between the actual user (generator of the intent) and the executing agent (executor of the API call). These two entities cannot be considered one anymore because one of them is driven by real intent while the other by LLM’s autonomous instructions, vulnerable to threats like Prompt Injection \cite{owasp-llm-top10-2025}, Excessive Agency \cite{owasp-llm-top10-2025}, and System Prompt Leakage \cite{owasp-llm-top10-2025}. Additionally, multiagent clients impose a more drastic challenge because different agents can potentially share the same client id and credentials leading to privilege escalations. 

To address these issues we propose a JWT extension protocol, based on both single and dual token design, that cryptographically encodes the user intent, delegation rules and separate identities for individual agents in a multiagent system. This protocol includes two new JWT token extensions, that allow the token to cryptographically encode intent and delegation assertion (they can be combined into a single token or even embedded in plain old JWT) to ensure that agents may not execute secure operations unless they are the intent owner or have been explicitly delegated that intent via a signed assertion issued by IDP

This paper makes the following contributions:
\begin{enumerate}
    \item Token design.  A formal specification of intent and delegation tied together in a single JWT token that cryptographically binds
    \begin{itemize}
        \item each agent action to a verifiable IDP registered user intent. 
        \item each agent action to a workflow step in an IDP approved workflow.
    \end{itemize} 
    \item IDP (Authorization Server) design and IDP side capablities required to support intent tokens.
    \item Resource Server side verifiation middleware design, fully backward compatible with OAuth 2.0 and JWT.
    \item Integrity tiers.  A checksum-based client shim and an optional TEE attestation \cite{10.1007/978-3-031-16092-9_7} profile that detect in-process impersonation with < 2 ms overhead.
    \item Prototype and analysis.  A reference implementation evaluated on a multi-agent micro-service, blocking 100 \% of threat requests.
\end{enumerate}

The remainder of the paper is organised as follows: §2 examines the background; §3 presents the archietcture and token suite; §4 defines the threat model; §5 describes security anchors as a follow up to the threat model; §6 illustrates the experimental framework used for reproducing the modeled threats; §7 discusses design trade offs and limitations; §8 concludes.
\section{Background: Identity and access management today}\label{sec:background}

\subsection{Classical IAM Model}\label{subsec:classical-model}

When OAuth 2.0 Protocol was created, its primary purpose was to address fundamental problems with the traditional direct client - server authentication process. It introduced an authorization layer (Identity Providers or IDP) that would allow separation of the role of the client (3rd party application acting on behalf of a resource owner) from that of the resource owner (RFC 6749 §1 Introduction) \cite{rfc6749}. The key point about this separation was the recognition that there has to be a cryptographic isolation between the resource owner (or end user) and the client entity (3rd party application) acting on behalf of the Resource Owner to access a protected resource hosted by a resource server. This isolation matters because it allows the protocol to avoid sharing and storage of permanent resource owner credentials on the client (3rd party application). It also allows for more flexibility and granularity of control over access required by a specific client independently of any other clients. In other words any given client acting on behalf of the resource owner represents a limited and granular scope approved by the resource owner instead of representing the resource owner in entirety, and it can do so without requiring access to the resource owner credentials. \cite{rfc6749}

Most of the latest token based authentication and authorization methods are based on this protocol. For example the JWT token (RFC 7519, \cite{rfc7519}) is one of the implementations of OAuth 2.0 protocol designed to contain and transfer authorization claims in a compact way using the JSON format having them integrity protected using digital signature (JWS, RFC 7515 \cite{rfc7515}), or encrypted using JWE (RFC 7516 \cite{rfc7516}).

This current state of identity and access management for http based communication is based on the understanding that a client (3rd party application formally provided authorization grant  by a resource owner to request access token from an authorization server) accurately represents the intent of the resource owner. The idea is to verify who is accessing a protected resource and what action is being performed. This is fundamentally a delegation model where the resource owner has delegated a slice of its permissions (represented by scopes included in a token) over a protected resource to the client to act on its behalf to perform the authorized actions.

NIST SP 800-63‐3 aligns with this separation.  It states that “federation allows a subscriber to leverage one credential across multiple relying parties, reducing credential management risk” (SP 800-63 §4.1)\cite{nist_sp800_63}.  The standard further classifies assertions into three Federation Assurance Levels (FAL).  At FAL 1, a bearer assertion is sufficient—“any party in possession of the assertion can use it”—whereas FAL 2/3 require binding the assertion to a proof-of-possession key (SP 800-63C §4.2)\cite{nist_sp800_63c}.  OAuth 2.0 bearer tokens and unsigned JWT access tokens therefore sit at FAL 1: convenient, but vulnerable if an autonomous agent unlawfully obtains the token.

\subsection{Delegation Semantics}\label{subsec:delegation-semantics}

While OAuth 2.0 solved the credential-sharing problem, real-world systems still require finer delegation techniques that go beyond a single user-to-client grant.

\begin{itemize}
    \item \emph{Scope-limited consent.} \\
    In the Authorization Code flow a resource owner approves a set of scopes (e.g., Calendar.read, Mail.send).  Every access token is therefore a subset of the user’s total privileges, narrowing blast radius if that token leaks \cite{rfc6749}.  However scopes are predefined and static; they cannot express lineage (which agent invoked whom) or context (why this action occurred).

    \item \emph{Machine-to-machine access.} \\
    The Client-Credentials grant lets a non-human client obtain a token for its own identity.  This is widely used for backend microservices but breaks accountability when the same client houses multiple autonomous agents sharing one credential set.

    \item \emph{Actor chains via Token Exchange.} \\ 
    RFC 8693 introduces a structured “actor” (act) claim that records when one party acts on behalf of another, enabling chained delegation such as Agent B acting for Agent A acting for Alice \cite{rfc8693}.  Yet the specification leaves enforcement semantics to implementers; resource servers still see a bearer token and must parse nested JSON to discover the chain.

    \item \emph{Proof-of-possession enhancements.} \\
    Draft DPoP binds each request to a public key and an HTTP signature, reducing replay attacks but not solving the intent-vs-execution split because the request still represents a single principal \cite{draft_dpop_2024}.
    
    \item \emph{GNAP's richer delegation model.} \\
    The Grant Negotiation and Authorization Protocol (GNAP) draft proposes fine-grained, dynamically negotiated access rights and first-class “sub-grant” objects for onward delegation \cite{draft_gnap_2024}.  Although promising, GNAP is early-stage and lacks widespread deployment.
    
\end{itemize}

Current mechanisms can chain identities or prove possession, but they do not cryptographically tie each downstream agent's action to the original user intent at runtime.  Nor do they prevent co-resident (running the same client process) agents from presenting the same bearer token.  These limitations motivate the intent-token / delegation-assertion design presented in § 3

\subsection{Zero Trust Foundations}\label{subsec:zero-trust-foundations}
How does all of this align with Zero Trust principles? Zero Trust changes the question from “Is this request coming from inside the perimeter?” to “Is this request, \textbf{at this moment}, \textbf{from a subject} I can \textbf{continuously verify}?”

The U.S. National Institute of Standards and Technology (NIST) formalises the model as “never trust, always verify, assume breach” (SP 800-207 §3).  In practice this means:

\begin{itemize}
    \item Strong identity for every principal, human or workload.

    \item Least privileged, fine-grained, context-aware authorisation tied to each request.

    \item Continuous evaluation of trust signals such as device health, network zone, and behavioural baselines.
\end{itemize}

Public-cloud examples include Google's BeyondCorp architecture, which proxies every request through a policy engine that re-authorises on each call rather than relying on long-lived sessions \cite{beyondcorp_google}.  CISA's Zero Trust Maturity Model extends the idea with “Just-In-Time and Just-Enough” access, emphasizing tokens that are short-lived and scope-limited \cite{cisa_zt_model_2021}.
Bearer OAuth/JWT tokens at FAL-1 (Section 2.1) satisfy strong identity at issuance time yet violate the continuous principle: possession alone remains sufficient until expiry.  When an autonomous agent obtains such a token—or prompts a co-resident agent (agent running in the same client process) to reuse one—the resource server cannot distinguish authorised intent from unauthorised execution.  The remainder of this paper shows how intent-scoped tokens with intent and delegation information cryptographically restore Zero-Trust guarantees in agentic workflows.

\section{Architecture}\label{sec:architecture}
The secure delegation architecture presented in this section implements the cryptographic agent 
authentication and intent delegation concetps described in our patent application 19/315,486 \cite{goswami2025agentic}.

\subsection{System Overview and Logical Actors}\label{subsec:system-overview}

The \agenticjwt system starts by adopting canonical OAuth 2.0 roles \cite{rfc6749} and embedding them, one-for-one, into the logical components of a Zero-Trust Architecture (ZTA) reference model (Fig. 2, § 3 of NIST SP 800-207 \cite{nist800207}).

Two additional refinements are introduced: 
\begin{enumerate}
    \item \emph{Agentic OAuth Client.} \\
    The traditional monolithic client is decomposed into an LLM-based Orchestrator Agent A and one or more Delegate Agents B\textsubscript{1}…B\textsubscript{n}, each of which deserves its own identity and least-privilege token.
    
    \item \emph{Client-Side Shim Library.} \\
    A tamper-proof shim library executes inside every agent process.  It computes agents' checksums, derives the per-agent PoP key, injects agentic headers (§ 4.4), and enforces the invariant that no agent can mint or replay a token on behalf of another agent.

\end{enumerate}

The \textbf{TABLE~\eqref{tab:logicalactors}} shows this mapping with short description of each of the components.

\begin{table}[!t]
    \caption{\textbf{Logical Actors}}
    \label{tab:logicalactors}
    \begin{tabular}{|p{37.5pt}|p{37.5pt}|p{37.5pt}|p{90pt}|}
        \hline
        \textbf{Logical Component RFC 6749}&
        \textbf{ZTA Logical Component NIST 800 207}&
        \textbf{ZTA Trust Zone Classification}&
        \textbf{Description} \\
        \hline
        Resource Owner&
        Subject&
        Untrusted&
        The end user that owns a protected resource (services and data) \\
        \hline
        Agentic OAuth Client&
        System&
        Untrusted&
        A multi agent application acting on behalf of the Resource Owner. This application embeds an LLM based orchestration agent A and one or more delegate agents B\textsubscript{1}...B\textsubscript{n}. \\
        \hline
        Authorization Server / Identity Provider (IDP)&
        Policy Decision Point (PDP)&
        Trusted&
        Trusted Identity provider that issues access tokens. \\
        \hline
        Client Shim Library&
        Policy Enforcement Point (PDP)&
        Trusted&
        A tamper proof client library loaded into every agent process. This library makes sure that individual agents' identities cannot be impersonated. \\        
        \hline
        Resource Server&
        Policy Enforcement Point (PEP)&
        Trusted&
        Exposes API endpoints and hosts the Protected Resource owned by the Resource Owner. \\
        \hline
        Protected Resource&
        Enterprise Resource&
        Implicit Trust Zone&
        The protected resource owned by the Resource Owner (end user). According to ZTA concepts this resource could be a protected service or data. \\
        \hline        
    \end{tabular}
\end{table}

\subsection{Agentic Client Internals}\label{subsec:agentic-internals}
\begin{figure*}[!t]
  \centering
  % Replace the filename with your exported PDF/SVG
  \includegraphics[width=\textwidth]{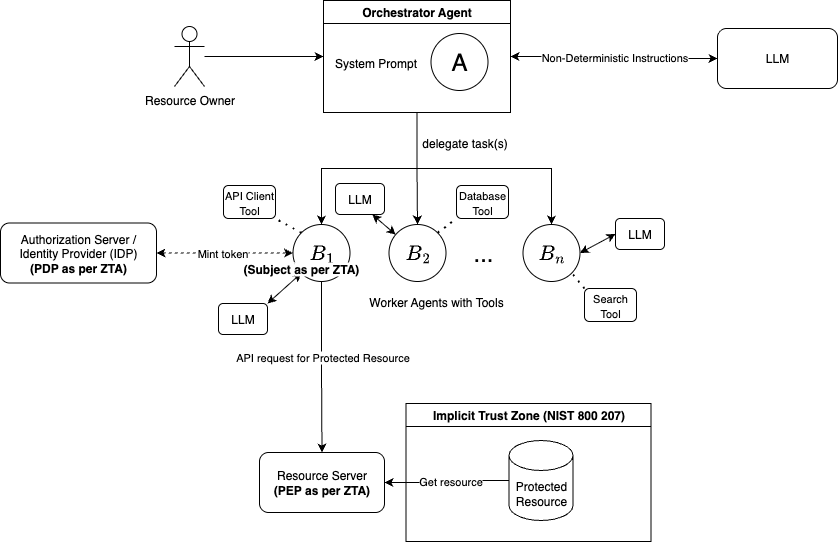}
  \caption{Agentic-client control flow.\@  Solid arrows denote data-plane
           calls; dashed arrows denote control-plane or token-minting calls.
           Components are mapped to their Zero-Trust roles
           (PDP, PEP, Subject, System) per NIST SP 800-207.}
  \label{fig:agentic-client-internals}
\end{figure*}

A typical end-to-end transaction inside an \emph{Agentic OAuth Client}
proceeds as follows (\textbf{FIGURE~\eqref{fig:agentic-client-internals}}):

\begin{enumerate}
    \item \textbf{User request} --- The \textit{Resource Owner} issues a task
        request (e.g.\ ``Show my VPN settings''), implicitly conveying a
        business intent.
    \item \textbf{LLM orchestration} --- The \textit{Orchestrator Agent~A}
        combines a system prompt with the request and calls the LLM.
        The LLM replies with a structured plan listing the delegate agents and
        arguments required to satisfy the intent.
    \item \textbf{Task delegation} --- Orchestrator~A spawns or signals the
        specified \textit{Delegate Agent(s)}~\(B_1\dots B_n\). 
    \item \textbf{Token minting} --- Whenever a delegate must call an external
        API, the shim authenticates to the \textit{Authorization Server} (Zero-Trust PDP)
        and receives a least-privilege access token (optionally bound to an
        \emph{intent token}). 
    \item \textbf{Shim library support} --- The process of interacting with IDP and Resource Server gets encapsulated 
        within the Shim library functionality. This includes the above \textbf{Token minting} step. 
        Each agent has access to this shim that does the following tasks: 
        \begin{enumerate}
            \item verifies the agent's separate identity registered with the IDP during authorzation grant,
            \item continuously tracks workflow that the multiagent client app is executing,
            \item computes agent's runtime checksum (based on prompt, tools and configuration parameters),
            \item mints intent token for any agent calling an API endpoint hosted on a Resource Server (agent is authecnticated based on its computed checksum and workflow delegation step executing at that moment),
            \item proves its own (Shim Library's) integrity via \texttt{X-Shim-Checksum}, 
            \item derives a PoP key, 
            \item injects agentic headers in the http request headers,
            \item sends the request to Resource Server,
            \item collects and returns response to the agent.
        \end{enumerate}
\end{enumerate}

\textbf{FIGURE~\eqref{fig:agentic-client-internals}} shows this flow with each of the components (along with Zero-Trust logical component labels). The Client Shim Library is installed such that its accessible to each of the Worker agents.

\subsection{Token Suite}\label{subsec:token-suite}
The OAuth 2.0 protocol perfectly aligns with Zero Trust Architecture (ZTA) 
tenets \cite{nist_sp800_207}. It organically supports the concepts of minimalistic implicit 
trust zones, continuous verification, dynamic context, least privileged access (using granular scopes) 
and services \& data as protected resource. The JWT token specification \cite{rfc7519} 
mentions that JWT is compact token encoded as JSON object \cite{rfc8259} in the payload of a 
JWS web signature \cite{rfc7515} or as plain text of a JWE \cite{rfc7519} structure that 
allows for a URL safe way of representing claims. Our Agentic JWT protocol builds on top of the 
OAuth and JWT specifications and proposes an extended set of tokens completely backward compatible 
with the OAuth and JWT specifications. The basic tenets of Agentic JWT align with those of ZTA 
and restore Zero Trust semantics in a scalable way when Agentic Clients attempt to access 
protected resources using OAuth and JWT.

The \agenticjwt system retains JSON token structure of JWT but brings the cryptographic authorization to the 
level of individual task intent instead of a user's scope. The intent of a task originates from 
the Resource Owner (end user) when it uses the OAuth Client Application (in this case the Agentic Client) 
to access a Protected Resource. A typical standard JWT uses scopes to model intent under assumptions 
of deterministic Client code. Scopes are used to represent authorization boundaries that control 
permissions and define what action(s) can a Client perform with the Protected Resource. If the client 
is a Multi Agent application controlled by an LLM, the number of combinations of different ways 
of performing API calls can explode undeterministically at runtime. It is not practical to 
predesign all the possible scope granularities and scale it to meet this possibility. Therefore, 
such Agentic Client systems with Multiple agents performing different types of actions, some 
requiring higher privileges, are forced to use a single coarse grained scope at the level of the 
client without differentiating one agent from another. 

In this scenario its not possible to use scopes to represent intent. We need to make the intent 
a first class citizen of the token protocol and provide a way to dynamically tie it to runtime 
identity of the caller Agent. We also need a way to cryptographically verify the intent on both, 
the Authorization Server / Identity Provider (IDP) and Resoruce Server sides. The scopes are not 
comparable to intent anymore because the actual execution of user intent is not deterministic. LLM 
powered Agents can access the Protected Resource based on LLM instructions which can change at runtime 
based on several factors such as Pompt Injection, Repeated Attempts, etc. leading to a separation 
between intent and execution. As an analogy, the scopes can be compared to having access into a building. 
But once you have access to the building you might have more specific tasks to perform, like 
organizing an event in the Auditorium, or Attending a meeting in Conference room no. 12. Some of 
these tasks could create lasting side effects, such as Renovating 5th floor to house a new 
office. Each of these specific tasks are implementation of some user's intent. With traditional 
non-agentic systems this intent used to be enforced automatically because it was implied from the 
deterministic and fixed client side code implementing a workflow. But in 
autonomous agentic systems the workflows are orchestrated by LLM's reasoning, it's like 
somehow these workflows are influenced by factors influencing LLM's reasoning and hence do not 
necessarily mirror a user's intent.

Philosophically, in the Agentic world, a scope grants potential; an intent grants permission 
for exactly one concrete task (or workflow) bounded by the scope boundaries.

Therefore, in the agentic world we are dealing with two fundamental security primitives that were 
previously hidden or naturally enforced, but are now exposed and need to be addressed: 
\begin{enumerate}
    \item \textbf{Identity} -- In the non-agentic world some fundamental properties were encoded 
    in and naturally enforced by fixed deterministic code. A deterministic workflow implemented 
    in a client app was easy to audit and never changed at runtime. There was a very clear 
    separation between data and instructions. Each of these fundamental properties made it natural 
    to define the Identity at the level of the client app. In the agentic world however, these 
    properties fade away because there is an autonomous decision taking LLM behind any action 
    taken by the agents \cite{yao2022react,schick2023toolformer,wang2023agentsurvey}. Therefore 
    agents (instead of the entire client app) form a more natural basis for building Identities. 
    We need to have separate identities for each agent. This concept is corroborated by several sources such 
    as W3C DID (identifier standard) \cite{w3c-did-2022}, W3C Verifiable Credentials 2.0 \cite{vc-data-model-2.0}, 
    and SPIFFE/SVID \cite{spiffe-concepts}. This approach directly supports Zero Trust for continuous 
    authorization around identities \cite{nist800207}
    \item \textbf{Intent} -- In the non-agentic world, deterministic code naturally enforced 
    fixed workflows and thus easy adherence to user's (Resource Owner) intent. There was no need 
    for directly binding credentials to intent or cryptographically verify it. These semantics do 
    not hold true anymore in case of agentic client apps. Hence, we need cryptographic verification 
    of intent. The internet statdards track specification RFC9396 establishes the need for more 
    granular information in JWT tokens specific scenarios \cite{rfc9396-rar}.
\end{enumerate}

Based on this reasoning the token suite is extended as follows: 

\subsubsection{Access Token}
This is an ordinary OAuth~2.0 bearer JWT as defined in \cite{rfc7519}.  
It represents the \emph{client application as a whole}, \emph{not} an individual
agent inside that application. For \textbf{deterministic} parts of the client 
application that \textbf{do not} use LLMs, the client can continue using the same 
old OAuth 2.0 compliant JWT tokens.

\begin{lstlisting}[label={lst:access-token}]
{
  "header": {
    "alg": "RS256",
    "typ": "JWT",
    "kid": "idp_key_2024"
  },
  "payload": {
    "iss": "https://idp.example.com",
    "sub": "vulnerability_scanner_v2.1",
    "aud": "api.github.com",
    "exp": 1719571200,
    "iat": 1719570900,
    "jti": "token_a7b9c2d4",
    "scope": "read:code write:report"    
  }
}
\end{lstlisting}

\subsubsection{Intent Token}
For non-deterministic parts that make use of LLM reasoning, each agent is supposed to have its 
own identity and permissions distinct from other agents even if they are running in the same 
Client process.

Each agent must obtain an \emph{Intent Token} issued separately from the client level access token, 
scoped to a \textbf{single agent + intent + workflow step}. This Intent Token is based on the running 
agent't identity which is computed by the Shim library described in sub-section 
\eqref{subsec:agentic-internals}.  Parsed as a standard JWT but with extra claims 
(`intent`, `agent\_proof`, `pop-jwk') understood only by servers that implement 
\agenticjwt. Resoruce Servers that do \textbf{not} implement the new \agenticjwt, can ignore 
these claims and the token should work as a regular JWT.

\begin{lstlisting}[label={lst:intent-token}]
{
  "header": {
    "alg": "RS256",
    "typ": "JWT",
    "kid": "idp_key_2024"
  },
  "payload": {
    "iss": "https://idp.example.com",
    "sub": "vulnerability_scanner_v2.1",
    "aud": "api.github.com",
    "exp": 1719571200,
    "iat": 1719570900,
    "jti": "token_a7b9c2d4",
    "scope": "read:code write:report",
    "intent": {
      "workflow_id": "vulnerability_assessment_v2",
      "workflow_step": "code_analysis",
      "executed_by": "static_analyzer",
      "initiated_by": "orchestrator",
      "delegation_chain": [
        "orchestrator",
        "static_analyzer"
      ],
      "step_sequence_hash": "sha256:1a2b3c4d",
      "execution_context": {
        "repository": "example/project",
        "branch": "main",
        "commit": "abc123"
      }
    },
    "agent_proof": {
      "agent_checksum": "sha256:a7b9c2d4e5f6...",
      "registration_id": "reg_1719570000",
      "version": "2.1.0"
    }
  }
}
\end{lstlisting} \cite{rfc7800}

\subsubsection{Delegation Assertion}
The 'intent' claim's 'workflow\_id', 'workflow\_step', and 'execution\_context' fields identify 
the current intent or workflow being executed. There is another aspect called delegation 
assertion which is represented by fields like 'initiated\_by', 'delegation\_chain', which tells 
the Resoruce Server about the delegation sequence that led to this API call.

\begin{lstlisting}[label={lst:delegation-assertion}]
{
  "header": {
    ...
  },
  "payload": {
    ...,
    "intent": {
      "workflow_id": "vulnerability_assessment_v2",
      "workflow_step": "code_analysis",
      "executed_by": "static_analyzer",
      "initiated_by": "orchestrator",
      "delegation_chain": [
        "orchestrator",
        "static_analyzer"
      ],
      "step_sequence_hash": "sha256:1a2b3c4d",
      "execution_context": {
        "repository": "example/project",
        "branch": "main",
        "commit": "abc123"
      }
    },
    ...
  }
}
\end{lstlisting}

If the IDP validates this assertion it returns the
corresponding Intent Token to the delegate.

\subsubsection{PoP Key}
Each agent generates (or is provisioned with) a short-lived
public key which is registered at the IDP; the private half is
used to sign the HTTP request (e.g.\ `Signature-Input`,
`Signature` headers, RFC 9440) \cite{rfc9440}.

\begin{lstlisting}
{
  "kty": "OKP",
  "crv": "Ed25519",
  "x":   "11qYAY...Sg",
  "kid": "agent:order-placer#2025-06-24T18:00Z"
}
\end{lstlisting}

The JWK’s thumb-print (`jkt`) is embedded in the `cnf` claim of
the Intent Token.

\subsubsection{X-Shim-Checksum}
Header applied to the client-side shim library (the
common enforcement layer) that allows IDP or Resource Server to validate the integrity of 
the Shim library running in the client environment.
\begin{lstlisting}{python}
X-Shim-Checksum: sha256 d72c55e9afe3...
\end{lstlisting}

\paragraph{\textbf{Why keep the schemas separate?}}
\begin{itemize}
    \item Access Token* remains fully interoperable; services that
  ignore agent-level semantics still work. The client applications may not be using LLMs 
  for every workflow step, these cases can use the plain old JWT access token. 
  LLMs would typically be used only for the cases where non-deterministic reasoning is 
  required to decide the next step to execute (possibly calling an API). 
    \item Intent Token* + PoP give fine-grained, cryptographically
  verifiable context to zero-trust-aware APIs without breaking
  legacy flows. The services that choose to ignore intent claims can still use this token 
  as a regular JWT.
\end{itemize}

\subsection{Reference Flows}

\subsubsection{Registration Flow \& Governance Model}

Typically the Agentic Client Registration is a one time process with some automated updates if required 
in case something changes on the client side. It can be carried out either manually via the User Interface 
provided by the IDP (Authorization Server), or can be automated via IDP CLI (Command Line Interface), 
or via the IDP APIs. The automated flow can be triggered using a standalone script or can be fully 
automated by invocation during the client application deployement from the CI / CD (Continuous Integration 
/ Continuous Deployment) pipeline execution. Agentic Client Registration involves registering 
LLM backed agents as separate identities to the IDP and maps under the Authorization Grant step 
of OAuth 2.0 process \cite{rfc6749}.

The Client Application is viewed as having an application level \textit{client\_id} with individual agents 
having their own ids (for example: \textit{agent\_A}, \textit{agent\_B}, etc.). This provides a clear 
separation of identity that can be used by the IDP to cryptographically isolate Agentic calls from non-agentic  
calls. The following Agentic Client Registration Flow facilitates this process. \linebreak

Participating Components: 
\begin{enumerate}
    \item Resource Owner
    \item Client Application \begin{enumerate}
        \item Non Agentic Part
        \item Agentic Part \begin{enumerate}
            \item Agent-A
            \item Agent-B
            \item Agent-C
        \end{enumerate}
        \item Client Shim
    \end{enumerate}
    \item Authorization Server (IDP) \\
\end{enumerate}

We will look at this Registration process by breaking it in two parts \textbf{1) Client Registration}, which registers an application at 
the level of a single Client, and \textbf{2) Agent Registration}, which registers each Agent in this Client app as a separate identity.
\begin{enumerate}
    \item Client Registration Sequence Flow: 
    This flow is described below. \textbf{FIGURE~\eqref{fig:client-registration-flow}} depicts the sequence steps of the proecess.
    \begin{figure*}[!t]
        \centering
        \includegraphics[width=\textwidth]{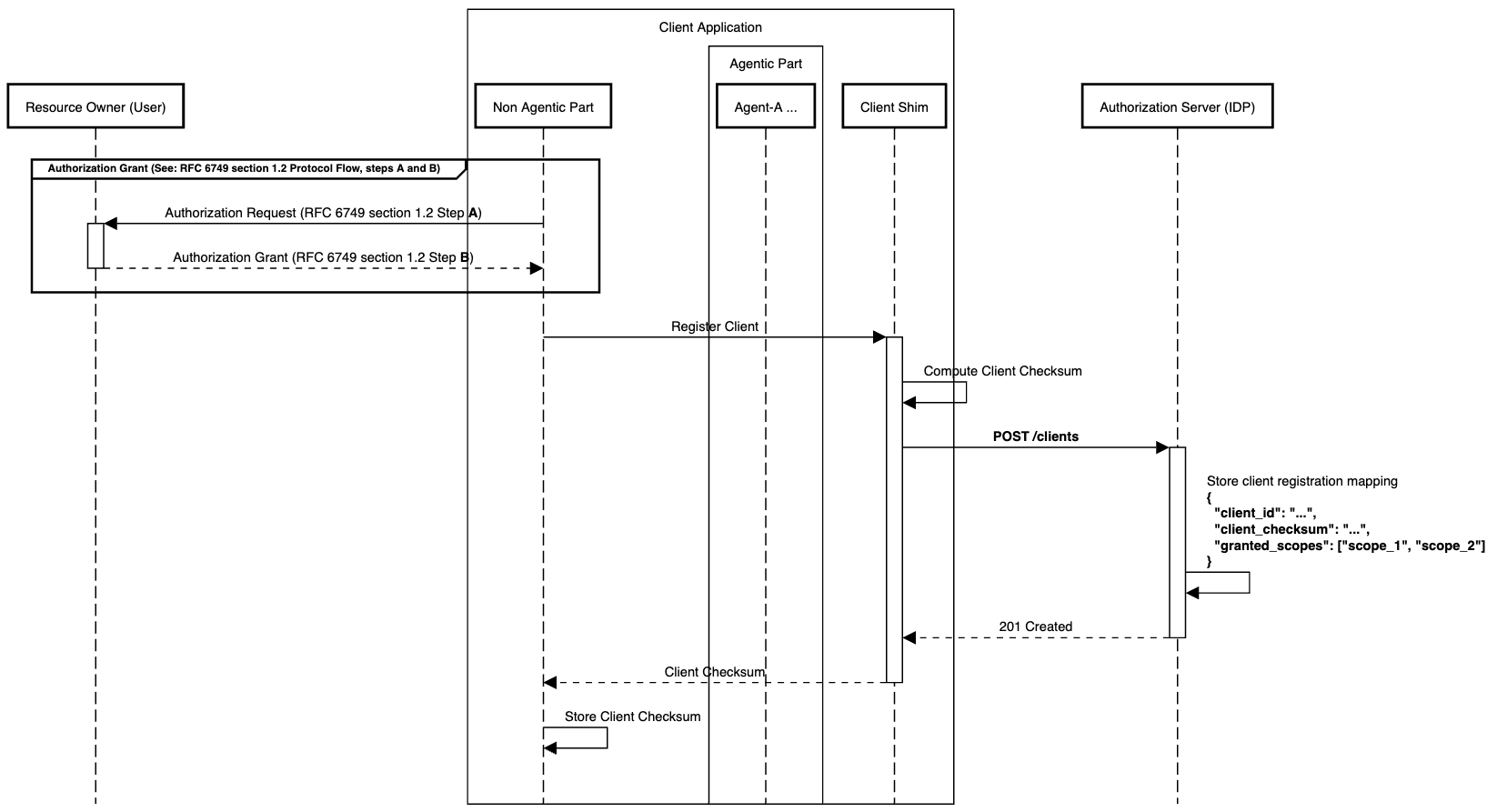}
        \captionsetup{justification=centering,singlelinecheck=false}
        \caption{\centering Client registration flow.}
        \label{fig:client-registration-flow}
    \end{figure*}
    \begin{enumerate}
        \item \textbf{Authorization Grant:} The client application obtains 'Authorization Grant' from the Resource Owner as outlined 
            in OAuth 2.0 specification, RFC 6749 section 1.2 \cite{rfc6749}. 
            \begin{itemize}
                \item Authorization Grant can be obtained either directly from the Resource Owner, or preferably indirectly 
                    via the Authorization Server. RFC 6749 section 1.2 \cite{rfc6749}
                \item Once the application Authorization Grant is obtained, it can be stored by the Client Application for 
                future. 
            \end{itemize}
        \item \textbf{Register Client as Agentic App:} The Client Application uses Authorization Grant obtained from the Resource Owner 
            (commonly via IDP), to request a registration.
            \begin{itemize}
                \item All the Client - IDP interactions take place via a specialized protocol aware client shim library 
                represented as 'Client Shim' in the sequence diagram.
                \item The 'Client Shim' library is backward compatible with OAuth 2.0 and only adds a simple step of 
                    computing Checksum of the client code before sending the registration request to the Authorization 
                    Server.
            \end{itemize}
        \item \textbf{Compute Client Checksum:} The request starts with being intercepted by the Client Shim library that computes a Checksum 
            for the Client Application. This Checksum is supposed to represent the binary identity of the 
            application being registered.
            \begin{itemize}
                \item Optionally for higher security environments, the Shim library is supposed to be tamper proof, 
                so that any attempts from the client to patch or change the interception and Checksum 
                computation behavior should be detected and should result in a decisive rejection of 
                the registration request.
                \item In case of manual registration the Client is responsible for providing the correct Checksum 
                    which will be stored during registration and verified on each token request.
            \end{itemize}    
        \item \textbf{POST /clients:} The Shim library sends and secure HTTP request to the Authorization Server for 
            registering the client with the IDP.
        \item \textbf{Store client registration mapping:} The IDP generates a client\_id and maps it to the provided 
            client checksum. It also verifies the provided Authorization Grant and grants the requested scopes to 
            the client. It stores a client record that represents mapping between this client\_id, client\_checksum 
            and a list of granted scopes.
        \item \textbf{201 Created} Authorization Server creates the client registration and returns back a 201 response.
        \item  \textbf{Client Checksum:} The Shim library returns Client Checksum 
    \end{enumerate}

    \item Agent Registration Sequence Flow: 
    The agent and workflow registration is described below. \textbf{FIGURE~\eqref{fig:agent-registration-flow}}
    depicts this process. The process can be automated or manual and is part of the Agent Governance and 
    Registration.
    \begin{figure*}[!t]
        \centering
        \includegraphics[width=\textwidth]{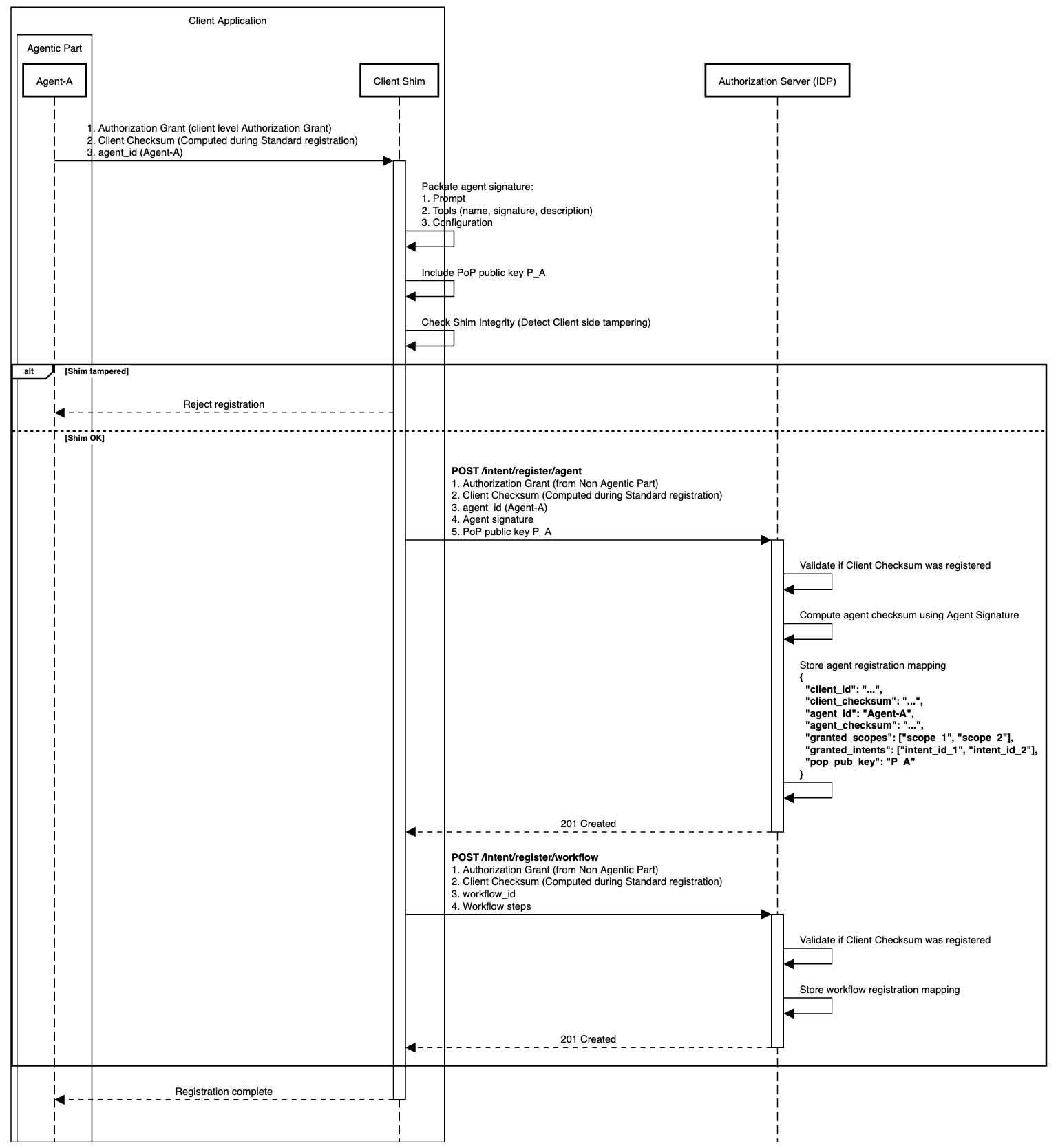}
        \captionsetup{justification=centering,singlelinecheck=false}
        \caption{\centering Agent registration flow.}
        \label{fig:agent-registration-flow}
    \end{figure*}
    \begin{enumerate}
        \item \textbf{Agent signature creation:} The Authoriation Grant and Client Checksum from the 
        Client Registration process are combined with the Agent signature to compute input 
        for Agent registration with IDP. Agent Signature is composed of: 
        \begin{itemize}
            \item Prompt: Agent prompt template and other prompt components.
            \item Tools: Tools that the agent uses. The name, signature and description of each tool.
            \item Configuration: Any agent or llm specific configuration parameters.
        \end{itemize}
        \item \textbf{Request to register Agent(s):} The request is sent to IDP for registering agent(s). This 
        request could be sent individually for each agent or in batch.
        \item \textbf{Request to register Workflow(s):} Separately from the agent registration workflows 
        can also be registered with IDP. Workflows are typically collection of steps taking place on an 
        agentic client application, each step can be imagined as the function or some executable that acts 
        as a tool used by an LLM backed agent. For e.g., if Agent-A as a tool in the form of a python function 
        that function can be considered a step in the Agentic workflow.
    \end{enumerate}
    At the end of the process the IDP ends up creating a record that consists of a client\_id / app\_id, 
    client checksum, agent\_id, agent checksum (computed on IDP), Agent's PoP public key sent in registration 
    request. It also separately creates records for each workflow registered by the client.

\end{enumerate}

\subsubsection[short]{Agentic Intent Token Minting}
\begin{figure*}[!t]
    \centering
    \includegraphics[width=\textwidth]{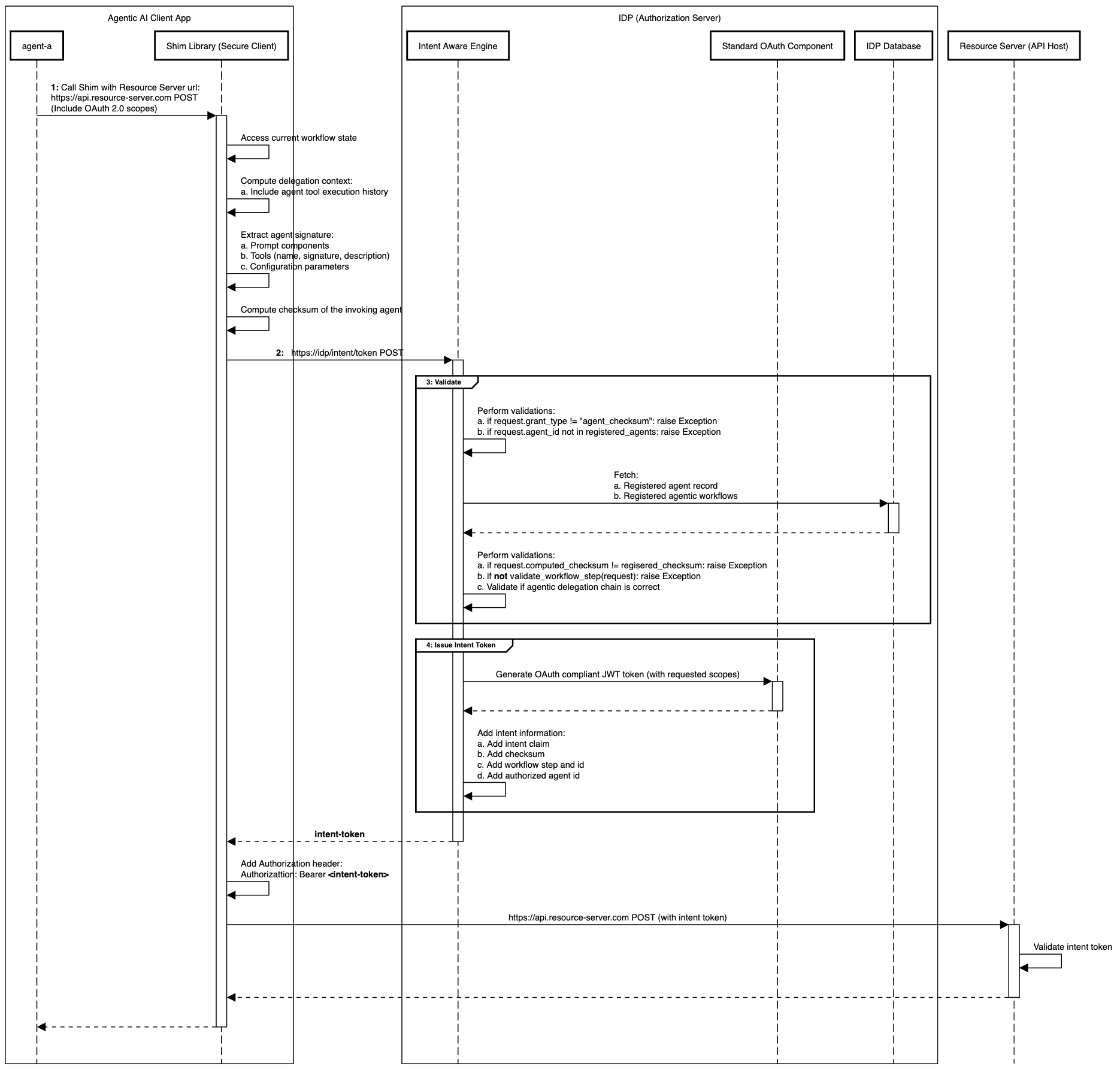}
    \captionsetup{justification=centering,singlelinecheck=false}
    \caption{\centering Agent token minting flow.}
    \label{fig:agent-token-minting-flow}
\end{figure*}

\textbf{FIGURE~\eqref{fig:agent-token-minting-flow}} illustrates the token minting flow.
\begin{enumerate}
    \item An agent when attempting to call the target API on Resource Server will go through the Shim 
    library, which will initiate token minting request with the IDP, this is just like the plain old 
    process of obtaining OAuth 2.0 access token (such as JWT token). This is shown as the Step 1 in the 
    diagram.
    \item The IDP after receiving the token request would typically perform a series of validations: 
        \begin{itemize}
            \item Verify that the agent exists and has been registered.
            \item Compare the provided checksum against the registered checksum of the agent which 
            amounts to runtime cryptographic check on agent's identity. 
            \item Validate workflow authoriziation for the requested workflow step.
            \item Check the delegation chain integrity
        \end{itemize}
    \item If successfully validated, the IDP issues and intent token (just like a traditional access token)
    containing identity and authorization information about the agent. The token as mentioned before 
    cryptographically binds an agents identity to the over intent represented by the current workflow 
    context.
\end{enumerate}

\subsubsection[short]{Agentic Workflow Tracking}
\begin{figure*}[!t]
    \centering
    \includegraphics[width=\textwidth]{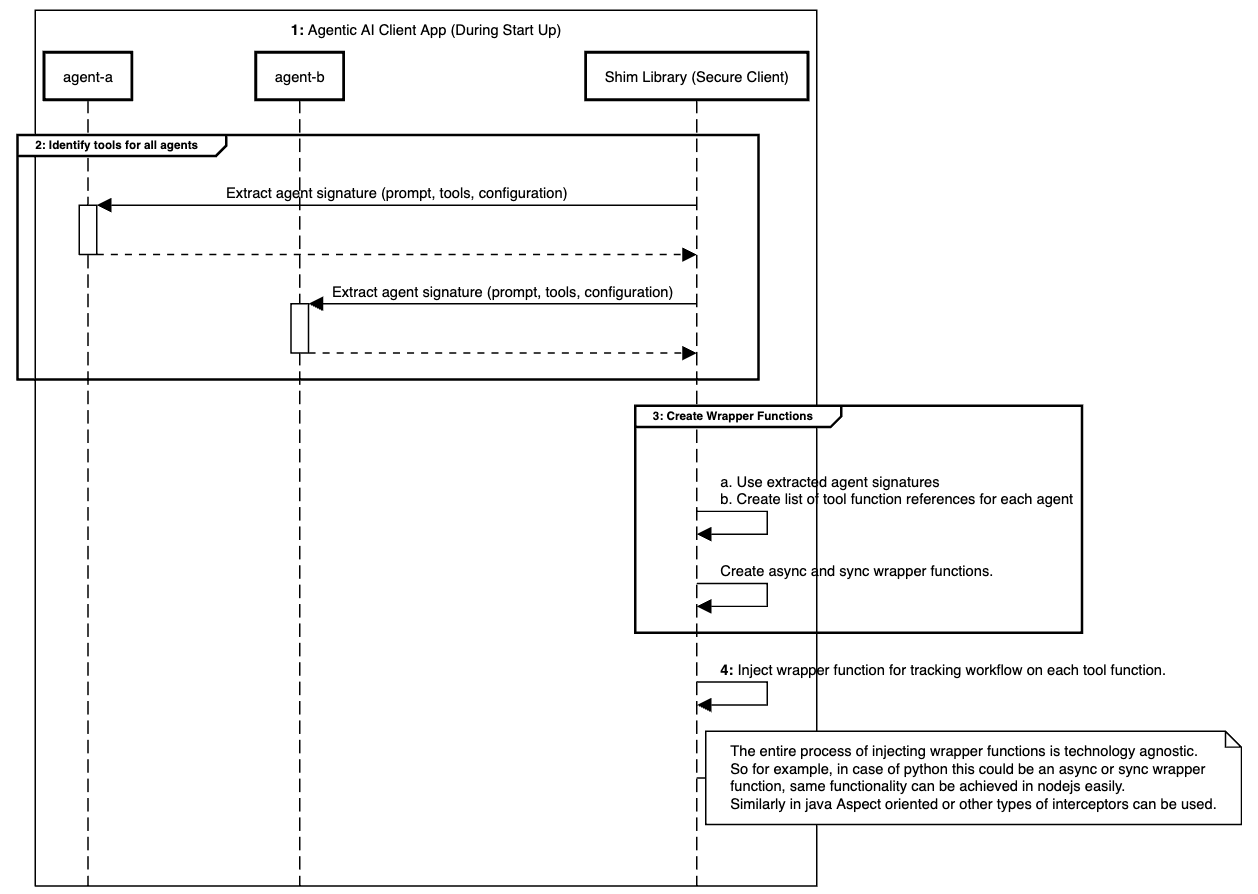}
    \captionsetup{justification=centering,singlelinecheck=false}
    \caption{Agentic Workflow Tracknig.}
    \label{fig:workflow-tracking-flow}
\end{figure*}

The Shim library is capable of agentic workflow tracking which means that as agents execute tool calling 
based on instructions from LLMs, the Shim library is able to track each tool call made as part of whatever 
workflow get executed by the agentic system and also able to capture the current state of the workflow at 
each tool call. If one ore more of these tool calls result in API calls to a Resource Server, that workflow 
state and entire agent delegation chain becomes part of the intent token minting request. 
\textbf{FIGURE~\eqref{fig:workflow-tracking-flow}} depicts this process in detail.

\section{Threat Model}\label{sec:threat-model}

\subsection{Methodology (STRIDE)}\label{subsec:stride-methodology}
We employ Microsoft STRIDE methodolgy for modeling threats and systematically analyze Agentic AI 
client applications for security threats. STRIDE provides comprehensive coverage across six threat 
categories, Spoofing, Tampering, Repudiation, Information Disclosure, Denial of Service, and Elevation 
of Privilege. Each identified threat is mapped to corresponding OWASP Top 10 vulnerabilities and 
real-world attack precedents to establish practical relevance.

The scope of our Threat Model includes the following system components, Resouce Owner, Agentic AI Client App, 
Client Shim Library, Authorization Server (Identity Provider or IDP), Resource Server (API host), and 
Protected Resource (hosted on the Resource Server). The system attempts to map to Zero-Trust architectual 
components as shown in the sub section \eqref{subsec:system-overview}. The goal is to restore Zero-Trust primitives 
and maintain them at all times while minimizing Implicit Trust Zones \cite{nist800207}. Our threat analysis 
recognizes that in case of Agentic AI Appliations we may need to draw Trust Boundaries between workflows 
running within the same client process, especially if those workflow are beign orchestrated by agents using 
LLM. This requires us to view each agent running within the same client process as a distict identity and 
a potential threat to other agents and workflows.
\subsection{Threat Enumeration}\label{subsec:threat-enumeration}
See \textbf{TABLE~\eqref{tab:threat-enumeration}} below for comprehensive threat list per STRIDE 
categories.
\begin{table*}[!t]
    \caption{\textbf{Threat Enumeration (STRIDE)}}
    \label{tab:threat-enumeration}
    \begin{tabular}{|p{45.25pt}|p{36.75pt}|>{\raggedright\arraybackslash}p{60pt}|>{\raggedright\arraybackslash}p{100pt}|>{\raggedright\arraybackslash}p{100pt}|>{\raggedright\arraybackslash}p{100pt}|}
        \hline
        \textbf{Threat Id}&
        \textbf{Category (STRIDE)}&
        \textbf{OWASP Mapping \cite{owasp-llm-top10-2025,owasp-top10-2021}}&
        \textbf{Description}&
        \textbf{Real-World Precedent}& 
        \textbf{Attack Vector} \\
        \hline
        T1:Agent Identity Spoofing&
        Spoofing&
        A01:2021 - Broken Access Control&
        A malicious agent impersonates a legitimate agent by replicating its code structure, prompts, and tool configurations to compute identical checksums&
        2019 Capital One breach where attackers used SSRF to assume IAM roles and access unauthorized resources. \cite{capital-one-occ-2020}& 
        Attacker gains access to agent source code (e.g., through repository compromise) and creates a malicious agent with identical checksum signatures. \\
        \hline
        T2:Token Replay Attacks&
        Spoofing&
        A02:2021 - Cryptographic Failures&
        Intercepted intent tokens are replayed by unauthorized agents to gain access to protected resources&
        2014 Heartbleed vulnerability led to OAuth token theft and unauthorized API access across major platforms. \cite{cve-heartbleed-2014}& 
        Network interception or memory dumps expose valid intent tokens that are replayed before expiration \\
        \hline
        T3:Shim Library Impersonation&
        Spoofing&
        1. LLM03:2025 - Supply Chain \newline \newline 2. A08:2021 - Software and Data Integrity Failures&
        Malicious replacement of legitimate shim library with compromised version that bypasses security controls&
        2020 SolarWinds Orion supply chain attack where legitimate software was replaced with backdoored versions. \cite{solarwinds-dhs-2021}& 
        Supply chain compromise or local privilege escalation to replace shim library files \\
        \hline
        T4:Runtime Code Modification&
        Tampering&
        A03:2021 - Injection&
        Agent prompts, tools, or configurations are modified at runtime after successful checksum registration&
        2021 vulnerability in certain versions of log4j java library where an attacker who could control 
        log messages or parameters could execute injected arbitrary code loaded from connected 
        LDAP servers during message lookup substitution. \cite{cve-2021-44228-log4j}& 
        Memory injection, debugger attachment, or reflection-based modification of agent properties \\
        \hline
        T5:Prompt Injection Attacks&
        Tampering&
        1. LLM01:2025 - Prompt Injection \newline \newline 2. LLM05:2025 - Inproper Output Handling&
        Malicious inputs cause LLM agents to generate unintended instructions that bypass security policies&
        2023 ChatGPT jailbreaking campaigns, Bing Chat prompt injection leading to unauthorized information disclosure. 
        Extensive study has been done on failures due to Jailbreak. \cite{wei2023jailbroken}& 
        Crafted user inputs or external data sources containing injection payloads that manipulate agent reasoning \\
        \hline
        T6:Workflow Definition Tampering&
        Tampering&
        A04:2021 - Insecure Design&
        Unauthorized modification of workflow definitions in the IDP to permit unauthorized agent transitions&
        CVE-2020-15228 \cite{cve-2020-15228-jenkins}. & 
        Compromised administrative credentials or IDP vulnerabilities allowing workflow redefinition \\
        \hline
        T7:Cross-Agent Privilege Escalation&
        Elevation of Privilege&
        1. LLM06:2025 - Excessive Agency, \newline \newline 2. A01:2021 - Broken Access Control&
        Lower-privilege agent manipulates higher-privilege agent to perform unauthorized operations beyond the original user intent&
        2020 Twitter OAuth tokens compromise allowing attackers to post from high-profile accounts \cite{twitter-hack-2020-doj}& 
        Agent A with read-only permissions crafts requests that cause Agent B with write permissions to execute destructive operations \\
        \hline
        T8:Workflow Step Bypass&
        Elevation of Privilege&
        A01:2021 - Broken Access Control&
        Agents skip required approval steps or execute workflow steps out of sequence to gain unauthorized access&
        Business logic flaws in e-commerce platforms allowing payment bypass. OWASP Guide \cite{owasp-business-logic-2020}& 
        Direct API calls bypassing workflow engine or manipulation of workflow state tracking \\
        \hline
        T9:Scope Inflation&
        Elevation of Privilege&
        LLM06:2025 - Excessive Agency&
        Agents request or utilize broader scopes than originally intended for the specific workflow step&
        OAuth scope creep vulnerabilities where applications request excessive permissions. RFC 6819 
        describes use of another client's crendentials for minting tokens \cite{rfc6819-oauth-security}& 
        Token minting requests with inflated scopes or misuse of broad scopes for unintended operations \\
        \hline
        T10:Intent Origin Forgery&
        Repudiation&
        A09:2021 - Security Logging and Monitoring Failures&
        Unable to cryptographically prove which user intent led to specific agent actions, enabling plausible deniability&
        Insufficient audit trails in financial systems leading to undetectable fraudulent transactions. Importance of 
        auduting systems discuss in NIST 800 92 \cite{nist-sp800-92-2006}& 
        Lack of cryptographic binding between user intent and downstream agent actions \\
        \hline
        T11:Delegation Chain Manipulation&
        Repudiation&
        A02:2021 - Cryptographic Failures&
        Modification or forgery of delegation chains to hide the true origin of agent actions&
        Certificate chain attacks in PKI systems allowing impersonation \cite{georgiev2012most}& 
        Manipulation of delegation assertion claims or replay of valid delegation chains in unauthorized contexts \\
        \hline
        T12:Agent Configuration Exposure&
        Information Disclosure&
        A01:2021 - Broken Access Control&
        Unauthorized access to agent prompts, tools, and configurations leading to system knowledge disclosure&
        2023 ChatGPT system prompt extractions revealing internal instructions. \cite{ignore_previous_prompt}& 
        API endpoints exposing agent metadata or memory dumps revealing agent configurations \\
        \hline                
    \end{tabular}
\end{table*}

\section{Security Anchors}\label{sec:security-anchors}
Security Anchors are fundamental design principles that address and directly mitigate each of the 
threats modeled in \eqref{subsec:threat-enumeration}, \textbf{TABLE~\eqref{tab:threat-enumeration}}. The new 
security protocol this paper proposes covers Authorization Server (Identity Provider or IDP), Client Shim 
Library, and Resource Server middleware (for verification of tokens). Therefore the design 
choices highlighted here would span across these components. Each of these components also 
maps to relevant Zero-Trust primitives / components (Please see \textbf{TABLE~\eqref{tab:logicalactors}}). 

\subsection{Identity Assurance Anchors}\label{subsubsec:identity-assurance-anchors}
\textbf{A1: Agent Checksum Verification} \newline
\textbf{Mitigates:} T1, T4, T12 \newline
\textbf{Mechanism:} The Shim library makes sure that the client application starts with agents 
that exactly match the registered agents including their id and checksums. At token minting 
time the Shim library computes runtime checksum of the calling agent and uses it to send 
intent token issuing request to IDP. The IDP verifies this computed checksum with registered 
checksum before issuing any tokens. The IDP does not allow registering agents with duplicate 
checksums. These principles mitigate T1 and T4.
The IDP includes checksum (which is a one way hash) of a given agent in the issued token. It 
does not include any plain text prompt, tools, or configuration (components that are used 
to compute the checksum, this directly addresses T12. \newline \newline
\textbf{A2: Registration First Security Model} \newline
\textbf{Mitigates:} T1, T3 \newline
\textbf{Mechanism:} All agent identities must be pre-registered with IDP during secure 
deployment process before any runtime operations. The IDP keeps a mapping between officially 
released Shim library versions and their binary checksums. IDP shares this with RS using a 
new /.well-known endpoint. The client side installation of Shim library always includes the 
actual runtime checksum in every interaction with IDP and RS which can be verified by them 
to detect tampering.
\textbf{Implementation:} CI/CD integration requiring cryptographic proof of legitimate 
deployment pipeline Or Manual Or Semi-Automated Governance process based on callng IDP 
provided admins console or CLI (Command Line Interface).
\textbf{Guarantee:} Unknown or unregistered agents cannot obtain valid tokens regardless 
of checksum knowledge. \newline \newline
\textbf{A3: Bridge Identifier Binding} \newline
\textbf{Mitigates:} T1, T7 \newline
\textbf{Mechanism:} Runtime agent detection based on unforgeable execution context 
(class references, memory layout). \newline
\textbf{Implementation:} Stack inspection and object identity verification at token 
request time. \newline
\textbf{Guarantee:} Agents cannot impersonate other agents within the same process space.

\subsection{Integrity Assurance Anchors}\label{subsubsec:integrity-assurance-anchors}
\textbf{A4: Client Credentials Prerequisite} \newline
\textbf{Mitigates:} T1, T2 \newline
\textbf{Mechanism:} In addition to the agent's runtime checksum, all inten token requests to 
IDP must include either the client application level credentials (normal OAuth authorization 
grant) or a standard JWT type of access token obtained at the client level deterministically 
using those credentials. \newline
\textbf{Implementation:} This becomes a two factor authentication which tells the IDP 
\textit{which agent}, running in \textit{which client application} is requesting an intent 
token. \newline
\textbf{Guarantee:} External attackers cannot mint tokens even if they gain knowledge of agent 
checksums. \newline \newline
\textbf{A5: Shim Library Integrity} \newline
\textbf{Mitigates:} T3, T4 \newline
\textbf{Mechanism:} Optional cryptographic validation of Shin library through 
self-attestation. \newline
\textbf{Implementation:} X-Shim-Checksum header in https headers can be verified by IDP 
as the IDP maintains a mapping of every Shim library version with its released checksum 
\newline
\textbf{Guarantee:} Client side Shim library cannot be updated or modified to change 
behavior. \newline \newline
\textbf{A6: Proof of Possession} \newline
\textbf{Mitigates:} T2, T11 \newline
\textbf{Mechanism:} During registration each agent generates ephemeral key pair and 
includes the public key in IDP registration request. The IDP stores it with this agent's 
registration record and sends in the cnf claim of the issued token. \newline
\textbf{Implementation:} Client performs Ed25519 signature with http request to the 
Resource Server. The Resource Server uses agent specific public key in the token and verifies
this signature. \newline
\textbf{Guarantee:} Intercepted or stolen tokens cannot be replayed without access to the 
agent specific private key. 

\subsection{Authorization Assurance Anchors}\label{subsubsec:authorization-assurance-anchors}
\textbf{A7: Cryptographic Intent Token Binding} \newline
\textbf{Mitigates:} T7, T9 \newline
\textbf{Mechanism:} Each API call is cryptographically bound to a specific intent (workflow) 
and a specific workflow step. \newline
\textbf{Implementation:} The \textit{Intent Token} issued by IDP contains immutable claims 
such as workflow\_id, workflow\_step, and execution\_context. These claims can be optionally 
encrpypted. \newline
\textbf{Guarantee:} Agents cannot execute actions or operations outside the approved workflows. 
These workflows are a representation of user intent. \newline \newline
\textbf{A8: Workflow Validation} \newline
\textbf{Mitigates:} T8, T9 \newline
\textbf{Mechanism:} Directed Acyclic Graph based workflow tracking and registration, insuring 
cryptographic verifiability of any arbitrary workflow. \newline
\textbf{Implementation:} IDP verifies curently running workflow step against registered and 
approved workflows. \newline
\textbf{Guarantee:} Agent's cannot drift from approved workflows. Workflow integrity is maintained 
if in the presence of factors that could influence LLM output. 

\subsection{Accountability Assurance Achors}\label{subsubsec:accountability-assurance-anchors}
\textbf{A9: Cryptographic Delegation Chains} \newline
\textbf{Mitigates:} T10, T11 \newline
\textbf{Mechanism:} The token contains delegation chain that represents an immutable audit train 
from user intent to the final agent in the workflow. \newline
\textbf{Implementation:} Delegation chains embedded in JWT claims with HMAC integrity 
protection. \newline
\textbf{Guarantee:} Complete provenance of every agent action can be cryptographically verified. 
\newline \newline
\textbf{A10: Workflow Execution Logging} \newline
\textbf{Mitigates:} T8, T10 \newline
\textbf{Mechanism:} Immutable logging of all workflow state transitions with timestamps. \newline
\textbf{Implementation:} Append-only logs with integrity protection for workflow events. \newline
\textbf{Guarantee:} All workflow deviations and workflow step changes can be recorded on IDP. 
\newline \newline
\textbf{A11: Agent Registration Versioning} \newline
\textbf{Mitigates:} T4, T6 \newline
\textbf{Mechanism:} All agent and workflow changes are versioned and create registration records on IDP. \newline
\textbf{Implementation:} Versioning system resides in IDP. \newline
\textbf{Guarantee:} All changes to agent behavior or workflow defintions are tracked and versioned. 
\newline \newline
\textbf{A12: Prompt Integrity Validation} \newline
\textbf{Mitigates:} T4, T5 \newline
\textbf{Mechanism:} Template-based checksumming with dynamic content validation. \newline
\textbf{Implementation:} Static prompt template hash verification plus whitelisted substitution rules. \newline
\textbf{Guarantee:} Prompt structure remains unchanged while allowing legitimate variable substitution. 
\newline \newline

\section{Experimental Framework}\label{sec:experimental-framework}
\subsection{Multi Agent Vulnerability Patching System}\label{subsec:patchet}
We created a simple but illustrative multi agent vulnerability patcher system 
that allows for reproduction of each of the threats highlighted in 
\eqref{subsec:threat-enumeration}. The goal is to monitor a git repository for 
push changes and trigger an agentic application that performs the following: 
\begin{itemize}
    \item Scan the file content of the repository and figure out the manifest 
    files (such as pom.xml, package.json etc.) used for dependency management 
    and upgrades, in a technology agnostic manner.
    \item Classify the manifest to find out the osv.dev ecosystem(s) that apply.
    \item Generate SBOM (software bill of material). Find out all the 
    packages and fetch their vulnerabilities from osv.dev.
    \item Triage the vulnerabilities and analyze them to create priority and plan 
    for patching each of them. 
    \item Group the vulnerabilities based on optimal way of creating Pull Requests. 
    \item Perform patching, create Pull Request, and Merge.
\end{itemize}

To achieve this goal the agentic application employs: 
\begin{enumerate}
    \item A Supervisor agent: that controls all decision and routing and uses other 
    agents as its tools. Does not call any APIs.
    \item A Planner agent: responsible for create that triaging, fetching vulnerabilities, 
    prioritizing and planning the patch operations. Calls github API to read repo manifests 
    and osv.dev API to fetch and analyze vulnerabilities.
    \item A Classifier agent: responsible for ecosystem classification. Calls osv.dev 
    API to fetch allowed ecosystem information.
    \item A Patcher agent: responsible for actual patching. Calls github POST calls or 
    to perform patching, create PR, and perform automatic merge. Requires Write permissions.
\end{enumerate}

\begin{figure*}[!t]
    \centering
    \includegraphics[width=\textwidth]{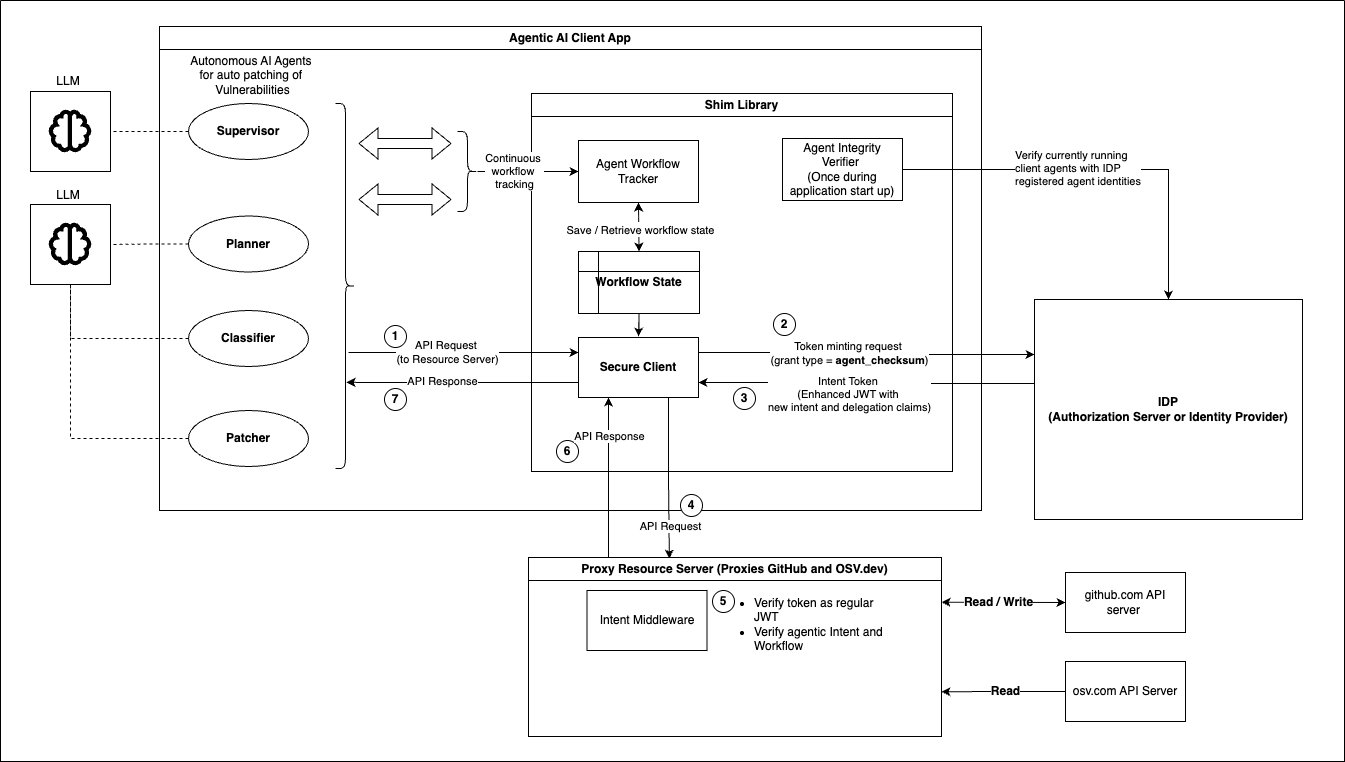}
    \captionsetup{justification=centering,singlelinecheck=false}
    \caption{Auto Patching Agentic App.\@ Figure adapted from Author's U.S. Patent Application No. 19/315,486, 2025 \cite{goswami2025agentic}.}
    \label{fig:auto-patching-app}
\end{figure*}

\textbf{FIGURE~\eqref{fig:auto-patching-app}} depicts this setup integrated within the Shim Library, IDP and 
a Resource Server Proxy that manages communication with github and osv.dev. In the 
real world the expectation is that API servers like github would directly support this 
new protocol so the Proxy Resource Server would not be required.

\subsection{Methodology}\label{subsec:experimental-methodology}
We have divided this experiment into a Before and After phase. The before phase uses normal 
OAuth 2.0 JWT tokens with authorization grant as 'client\_credentials'. The After phase 
uses \agenticjwt based \textit{Intent Token} with authorization grant as 'agent\_checksum'

We reproduce the threats outlined in \eqref{subsec:threat-enumeration} and observe that these 
threats exist in the traditional JWT token environment.
We then run the exact same scenarios in After phase with intent tokens and observe that the 
threats are mitigated by denying token minting or denying resouce server access, or other 
security enforcement methods.

\textbf{Note:} Full experimental results and performance analysis would be a part of our 
forthcoming journal publication

\section{Design Trade-offs \& Limitations}\label{sec:design-tradeoffs}
Enterprise level Agentic AI applications would almost always use APIs for performing 
actions. Therefore, APIs are the chokepoints where agents actually affect the real 
world and it makes sense to secure API endpoints. The current \agenticjwt system is 
a way to achieve this, but as with all designs it does have some tradeoffs. Below is 
an enumeration of some major tradeoffs.

\subsection{Performance and Scalability}\label{subsec:perf-tradeoff}

\subsubsection{Token Minting Latency}\label{subsubsec:tml} The use of short-lived or one-time tokens can increase the average 
frequency of minting tokens. The Shim library reference implementation does provide a token caching 
mechanism but it's user depends on the security policy and configured token expiry.

\subsubsection{Workflow Introspection during token minting}\label{subsubsec:workflow-introspection} 
The workflow introspection does not require any external call and is completely in-memory it 
may add to latency, although the impact is likely to be minimal.

\subsubsection{Scaling the workflow registration}\label{subsubsec:scaling-workflow} Any time a 
new workflow is designed in an Agentic App it will have to be registerd with the IDP. This 
practice poses some scalability challenges and makes the governance more complex than the 
standard OAuth 2.0 model. There can be some logical ways to mitigate this, for example the 
system allow disabling of workflow tracking, but this comes at the cost of not mitigating 
some of the threats outlined in \eqref{sec:threat-model}. Alternatively, the system can 
provide governance automation during offline time via scripts or during CI / CD process to 
infer workflows from the source code and register with IDP. This automation is not currently 
part of the reference implementation due to the development complexity involved in achieving 
it in a technology agnostic manner.

\subsection{Deployment Complexity}\label{subsec:deployment-tradeoff}

\subsubsection{IDP}\label{subsubsec:idp} As compared to IDP implementations supporting standard OAuth 2.0, the IDP issuing 
intent tokens with workflow delegation chains is more complex and specialized. The reference 
implementation does provide a good base to build on.

\subsubsection{Workflow and Agent Regisration}\label{subsubsec:workflow-reg} The agents need to register every time their signature 
(prompt, tools or configuration) changes. This means there has to be some Governance layer that 
allows for automation to simplify this process. For example, if prompts, tools etc. are managed 
using MCP (Model, Context, Protocol) servers, the IDP registration on every prompt update will require integrating with 
MCP.

\subsection{Technical Constraints}\label{subsec:technical-constraints}
\subsubsection{Runtime Agent Identity}\label{subsubsec:runtine-identity} When an agent runs a tool 
that invokes an API, the Shim library during token minting computes the runtime checksum of the 
running agent which it does by using a technique called \textit{Bridge Identifier}. What it means 
is that we consider a program specific anchor like a class or a function that would be present 
during start up (when client side agent cheksums are computed) and also present during the agent's 
execution. During the execution or token minting time the systems extracts the computed checksum 
using this bridge identifier. This concept is full proof but may need language specific implementation 
in some cases.

\subsubsection{Reference Implementation language specific}\label{subsubsec:ref-implementation-lang-specific }
The reference implementation, although technologi agnostic in general concept but is specific to 
python. For other language it would need to be implemented separately. This is not really a 
limitation but more of a tradeoff which is applicable to almost any analogous software.

\subsection{Security TradeOffs}\label{subsec:security-tradeoffs}
\subsubsection{TOCTOU Gap in Prompt Evaluation}\label{subsubsec:toctou}
Any agent's checksum identity is based on its prompt, tools and configuration. To me precise, 
the prompt part is mostly prompt template with some metainformation about variables that may 
get replaced on runtime. In any meaningfull agentic app the actual prompt will have placeholders 
which will get substituted using runtime values coming from either the user input or previous 
tool call outputs. Now the change in prompt due to these template substitutions, is difficult to 
distinguish from prompt injections. The system handles this in a way so as to allow genuine 
dynamic changes while rejecting prompt inejction cases, but this is yet to be tested across a 
range of real world situations.

\subsubsection{Information Disclosure Trade-Off}\label{subsubsec:information-disclosure}
Including workflow metadata in intent tokens enables fine-grained authorization but potentially exposes workflow details. Organizations can mitigate this through:
\begin{itemize}
    \item Application-specific encryption of workflow fields. The Intent IDP supports this by allowing
    app specific asymmetric key pair generation and associating it with registerd workflow record.
    \item Opaque workflow identifiers like arbitrary labels mapped to actual workflow step names. This mapping can be 
    registered with IDP. The reference implementation can be easily enhanced to support this out-of-the-box.
    \item Risk acceptance based on operational requirements.
\end{itemize}

\subsubsection{Additional Cryptographic Complexity}\label{subsubsec:additional-cc}
PoP Keys, Workflow encryption etc. introduce more complexity overall and thus require a Shim 
library for automation and support. The Shim library can be easily installed by adding it as a 
simple dependency in any language specific build manifest, such as package.json, pom.xml etc.

\subsection{Adoption Barrier}\label{subsec:adoption-barrier}
\subsubsection{Ecosystem}\label{subsubsec:ecosystem}
For full effectiveness of this system, ecosystem coordination and standardization is required. 
For example, The Shim library needs to be adopted by Agentic client apps, the IDP by both Agentic 
clients and Resoruce Servers. Similarly the Resource Server side middleware that performs token 
verification needs to be adopted as well. This adoption may take some time to mature.

\subsubsection{Standardization}\label{subsubsec:standardization}
The new Intent Token protocol is built on existing OAuth 2.0 JWT implementation. It needs to be 
standardized via the IETF (Internet Engineering Task Force) standard, which is subject to a 
significant review process and approval.

\section*{Acknowledgment}
The author thanks Dr.\ Prasenjit Shil (Senior Member, IEEE) for valuable feedback and proofreading 
during the preparation of this manuscript.

This research was conducted independently by the author. The work, findings, methods, and conclusions are 
solely those of the author and do not necessarily reflect the views or positions of any affiliated institution 
or employer. Any use of institutional email addresses is for identification purposes only and does not imply 
sponsorship, funding, or endorsement.

\textbf{Funding:} No funding was received for this work.

\textbf{Competing interests:} The author declares no competing interests.

% ----- Bibliography -----
\bibliographystyle{IEEEtran}
\bibliography{refs}

% ----- (Optional) biography section -----
% Ensure biography.tex uses IEEEtran's \begin{IEEEbiography} ... \end{IEEEbiography}
\begin{IEEEbiographynophoto}{Abhishek Goswami} 
received the Bachelor of Engineering degree in computer science from Barkatullah University, India, 
and the M.B.A. degree from University of Chicago Booth School of Business, Chicago, IL, USA. He is 
currently a Strategic Industry Expert specializing in Data \& AI, Cybersecurity, and Cloud Architecture.

Abhishek has helped organizations across the globe innovate, design, and implement research-grade 
solutions to solve complex real-world problems. From 2005 to 2010, he was a multi-disciplinary 
practitioner of applied software engineering, solving challenges across logistics, retail, manufacturing, 
heavy engineering, turbine engineering, and automotive domains. From 2010 to 2019, he served in 
progressive roles including Lead Engineer, Engineering Manager, and Technology Architect for 
Cybersecurity, Artificial Intelligence, and Cloud projects in Finance, Public Services, and 
Taxation sectors. Since 2020, his work focuses on industry implementation of AI and Cybersecurity research, 
providing strategic technology guidance in Power \& Energy, Banking, and Healthcare domains.

He became a Member (M) of IEEE in 2019, a Senior Member (SM) in 2025, and was invited to become 
an Industry Professional Member of Eta Kappa Nu IEEE Honor Society in 2025. His research interests 
include Deep Generative Models, AI agent security, cryptographic authentication systems, 
AI governance frameworks, and enterprise \& cloud architecture.
\end{IEEEbiographynophoto}

\end{document}